\documentclass[letterpaper, 10 pt, conference]{ieeeconf}  

\IEEEoverridecommandlockouts                              

\usepackage{graphics} 
\usepackage{epsfig} 
\usepackage{mathptmx} 
\usepackage{times} 
\usepackage{amsmath} 
\usepackage{amssymb}  
\usepackage{cite}
\usepackage{xcolor}
\usepackage{soul}
\usepackage{hyperref}
\usepackage{fancyhdr}

\fancypagestyle{plain}{
  \fancyhf{}
  \fancyhead[C]{Conference on \LaTeX}     
  \fancyfoot[L]{This is a notice}

}
\usepackage{eso-pic}

\title{\LARGE \bf
 A Simplified 3D Ultrasound Freehand Imaging Framework Using 1D Linear Probe and Low-Cost Mechanical Track}

\author{Antony Jerald$^{1}$, Madhavanunni A. N.$^{1}$, Gayathri Malamal$^{1}$, Pisharody Harikrishnan Gopalakrishnan$^{1}$ \\ and Mahesh Raveendranatha Panicker$^{1}$
\thanks{$^{1}$Antony Jerald, Madhavanunni A. N., Gayathri Malamal, Pisharody Harikrishnan Gopalakrishnan and Mahesh Raveendranatha Panicker (Email: mahesh@iitpkd.ac.in) are with the Centre for Computational Imaging and Department of Electrical Engineering, Indian Institute of Technology Palakkad, Kerala, India}}

\begin{document}

\AddToShipoutPictureBG*{%
  \AtPageUpperLeft{%
    \setlength\unitlength{1in}%
    \hspace*{\dimexpr0.5\paperwidth\relax}
    \makebox(0,-0.75)[c]{\textcolor{red}{\large This is an originally submitted version and has not been reviewed by independent peers.}}%
}}

\AddToShipoutPictureBG*{%
  \AtPageLowerLeft{%
    \setlength\unitlength{1in}%
    \hspace*{\dimexpr0.5\paperwidth\relax}
    \makebox(0,0.75)[c]{\textcolor{red}{\textit{This work is licensed under a \href{https://creativecommons.org/licenses/by-nc-nd/4.0/}{Creative Commons Attribution-NonCommercial-NoDerivatives (CC-BY-NC-ND) 4.0 License.}}}}%
}}

\maketitle
\thispagestyle{empty}
\pagestyle{empty}

\begin{abstract}

Ultrasound imaging is the most popular medical imaging modality for point-of-care bedside imaging. However, 2D ultrasound imaging provides only limited views of the organ of interest, making diagnosis challenging. To overcome this, 3D ultrasound imaging was developed, which uses 2D ultrasound images and their orientation/position to reconstruct 3D volumes. The accurate position estimation of the ultrasound probe at low cost has always stood as a challenging task in 3D reconstruction. In this study, we propose a novel approach of using a mechanical track for ultrasound scanning, which restricts the probe motion to a linear plane, simplifying the acquisition and hence the reconstruction process. We also present an end-to-end pipeline for 3D ultrasound volume reconstruction and demonstrate its efficacy with an \textit{in-vitro} tube phantom study and an \textit{ex-vivo} bone experiment. The comparison between a sensorless freehand and the proposed mechanical track based acquisition is available online (shorturl.at/jqvX0).\newline


\indent \textit{Clinical relevance}— The proposed approach simplifies the process of 3D ultrasound acquisition and reconstruction and has the potential to enhance the diagnostic accuracy and precision of point-of-care ultrasound imaging.
\end{abstract}

\section{INTRODUCTION}
Ultrasound imaging is a popular point-of-care medical imaging modality due to its several advantages such as dynamic imaging, lack of ionizing radiations, comparatively low cost, and easy disinfection. However, conventional 2D ultrasound imaging suffers many disadvantages \cite{c1}. The most important is that the decision-making process in diagnosis and analysis is time-consuming and might lead to inaccurate judgments. This is because the physician needs to mentally transform a sequence of 2D ultrasound images to produce a 3D volume representation. Also, it is challenging to relocate the exact orientation of any previously captured image to record the progression and regression of pathology in response to therapy. Measurement of an organ volume, which is very important in some treatments and surgery, is highly inaccurate in 2D ultrasound imaging. Due to these limitations of 2D ultrasound imaging, there is a growing need for 3D ultrasound imaging, which provides more precise and integrated information about organs and is better suited for treatment and clinical diagnosis \cite{c14}. 

Various ultrasound data acquisition methods using 2D arrays and mechanical 3D probes have been developed over the years. Using a dedicated 2D array ultrasound probe is one of the fastest ways to view the 3D volume in real-time as it does not require precise orientation information for volume generation \cite{c16}. But a 2D array probe is costly and complex to design in terms of both hardware and software. Moreover the limited size of 2D array transducers, due to difficulties in fabrication, results in a smaller field of view in imaging \cite{c1}. In a mechanical 3D probe, a standard linear array transducer is motored to translate, rotate and tilt within the probe to acquire images, and the translation, rotation, and axis of rotation can be used as a reference for reconstruction. However, the mechanical 3D probes need to be held statically by the doctors while images are being taken, which introduces latent flaws in the data-collecting process \cite{c2}. Another method uses motorized mechanisms to tilt or translate the conventional linear array probe and rapidly acquire 2D ultrasound images \cite{c18}. Here the scanning position and orientation need to be predefined and are controlled by a stepper/servo motor, hence it acquires regularly spaced 2D ultrasound frames. However, these motorised probes are bulky, making them inconvenient for frequent scanning purposes \cite{c2}. 

In freehand ultrasound acquisition, orientation estimation sensors like electromagnetic (EM) tracking sensors are attached to the ultrasound probe to register the orientation/position of each frame along with the ultrasound frame. This method allows the user to move the probe freely, making it more flexible in terms of mobility \cite{c3}. However, EM trackers can be affected by interference from nearby magnetic sources, resulting in decreased tracking accuracy \cite{c1}. Deep learning-based sensor-less freehand ultrasound 3D reconstruction \cite{c20} techniques have gained popularity in the recent past. But the deep learning-based approach requires a lot of labelled training samples and high computational power for accurate results \cite{c19}. Also, in general, most of the approaches in literature have been proposed to enable 3D scanning on a moderately high-cost system not intended for point-of-care imaging.

In this work, a novel approach to overcome the challenge of orientation/position estimation in freehand-based scanning is attempted by using a low-cost mechanical track that simplifies position tracking using a predefined movement region. Interpolation techniques are utilized to reconstruct the acquired data into a 3D volume, which can then be visualized through volume rendering. 
\begin{figure*}
    \begin{center}
        \includegraphics[width= 0.90\linewidth]{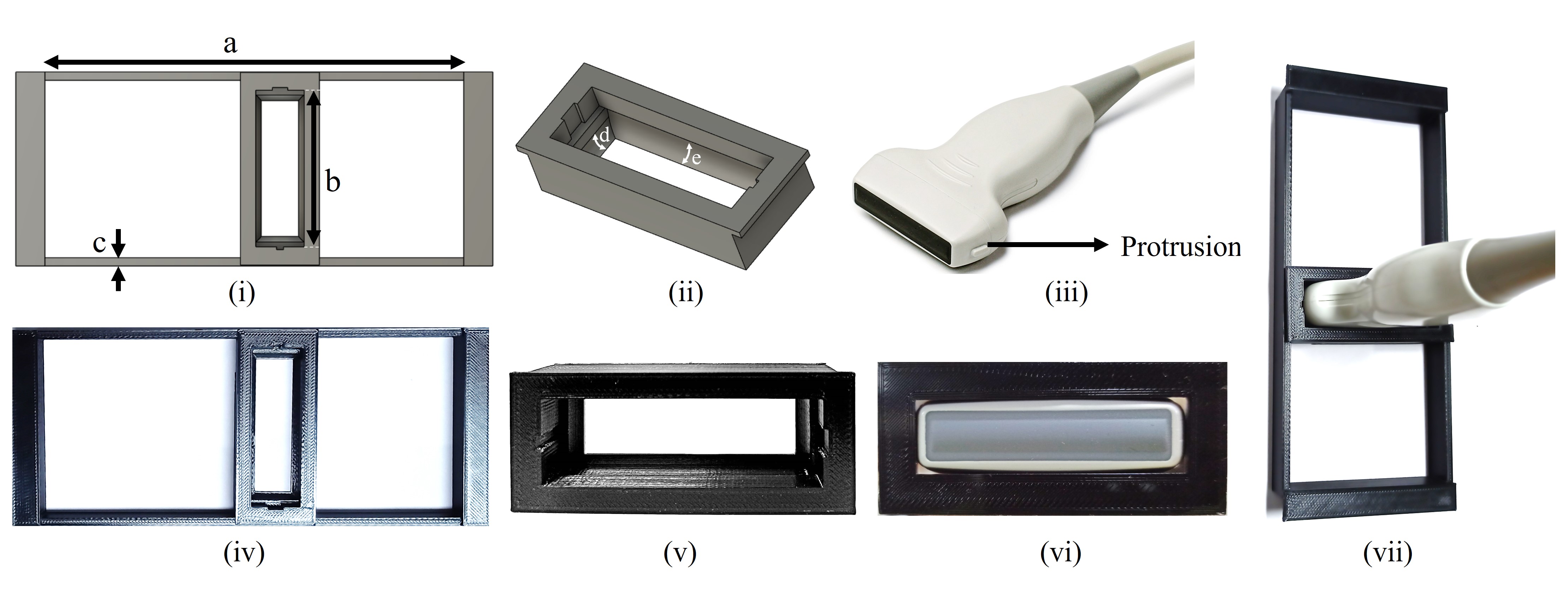}
        \caption{(i) and (ii) Rectangular track and mask designed in Fusion 360, (iii) Verasonics L11-5v probe, (iv) and (v) 3D printed track and mask, (vi) and (vii) Probe on track}
        \label{Track and Probe}
    \end{center}
\end{figure*}

\section{Proposed Approach}
\label{Proposed Approach}
\subsection{Proposed Mechanical Track}

The use of a mechanical track for probe motion in 3D acquisition simplifies position tracking by confining the probe to linear motion in a plane. The mechanical track consists of two components: the probe mask, which reduces errors in linear scanning by preventing the probe from tilting, and the rectangular track, which serves as a guide rail for the smooth movement of the probe and prevents slipping from the operator’s hands. The mechanical track, designed using Fusion 360 software (Fig. \ref{Track and Probe}(i) \& (ii)) to match the probe and scanning material, is 3D printed (Fig. \ref{Track and Probe}(iv) \& (v)) and is cost-effective. The Verasonics L11-5v probe (Fig. \ref{Track and Probe}(iii)) is used in the proposed study. The track width is selected to guarantee seamless mask mobility. A curved bottom of the mask matching the probe shape is designed to ensure a snug fit (Fig. \ref{Track and Probe}(vi) \& (vii)) and prevents the vertical falling of the probe. Side cuts on the mask are provided to accommodate the probe protrusions. Table \ref{tab:Rectangular Track Dimension} summarizes the dimensions of the designed track and mask. The frames are acquired at high frame rates employing multi-angle plane-wave compounding which results in a high correlation among the acquired frames. This leads to low-complex 3D volume reconstruction using simple trilinear interpolation between frames, simplifying the volume reconstruction process and reducing computational complexity.

\begin{table}[t!]
    \centering
	\caption{Mechanical Track Dimensions}
    \begin{tabular}{ccc}
    \hline
    \hline
    \textbf{Symbol}     &     \textbf{Dimension (cm)}    &     \textbf{Use} \\ 
    
       \hline
    a  & 12 &  Length for linear motion of probe  \\ 
    b  & 5.3 & Width of the probe  \\ 
    c  & 0.2 &  Overlap which acts as a railing  \\ 
    d & Curve with radius of 7 &  For fitting the curved bottom part \\ 
    e & Curve with radius of 14 &  For fitting the curved bottom part\\ 
    \hline
    \hline
    \end{tabular}
    \label{tab:Rectangular Track Dimension}
\end{table}

\AddToShipoutPictureBG*{%
  \AtPageUpperLeft{%
    \setlength\unitlength{1in}%
    \hspace*{\dimexpr0.5\paperwidth\relax}
    \makebox(0,-0.75)[c]{\textcolor{red}{\large This is an originally submitted version and has not been reviewed by independent peers.}}%
}}

\AddToShipoutPictureBG*{%
  \AtPageLowerLeft{%
    \setlength\unitlength{1in}%
    \hspace*{\dimexpr0.5\paperwidth\relax}
    \makebox(0,0.75)[c]{\textcolor{red}{\textit{This work is licensed under a \href{https://creativecommons.org/licenses/by-nc-nd/4.0/}{Creative Commons Attribution-NonCommercial-NoDerivatives (CC-BY-NC-ND) 4.0 License.}}}}%
}}

\subsection{Proposed Image Processing Pipeline}
Ultrasound images are typically characterized by speckles and hence the selection of image processing algorithms is a crucial stage in the reconstruction pipeline. The various steps in the proposed image processing pipeline are shown in Fig. \ref{Image Processing Steps} and explained in detail in the subsequent sections. 

\subsubsection{Log Compression and Squaring}
Log compression enhances the low-intensity values in images with strong reflections. Squaring the image increases contrast by intensifying the dark areas and brightening the bright regions. The improvement in the visibility of the acquired ultrasound images is illustrated in Fig. \ref{Image Processing Steps} (b).

\subsubsection{Median filtering}
\begin{figure*}
    \begin{center}
        \includegraphics[width= 0.95 \textwidth]{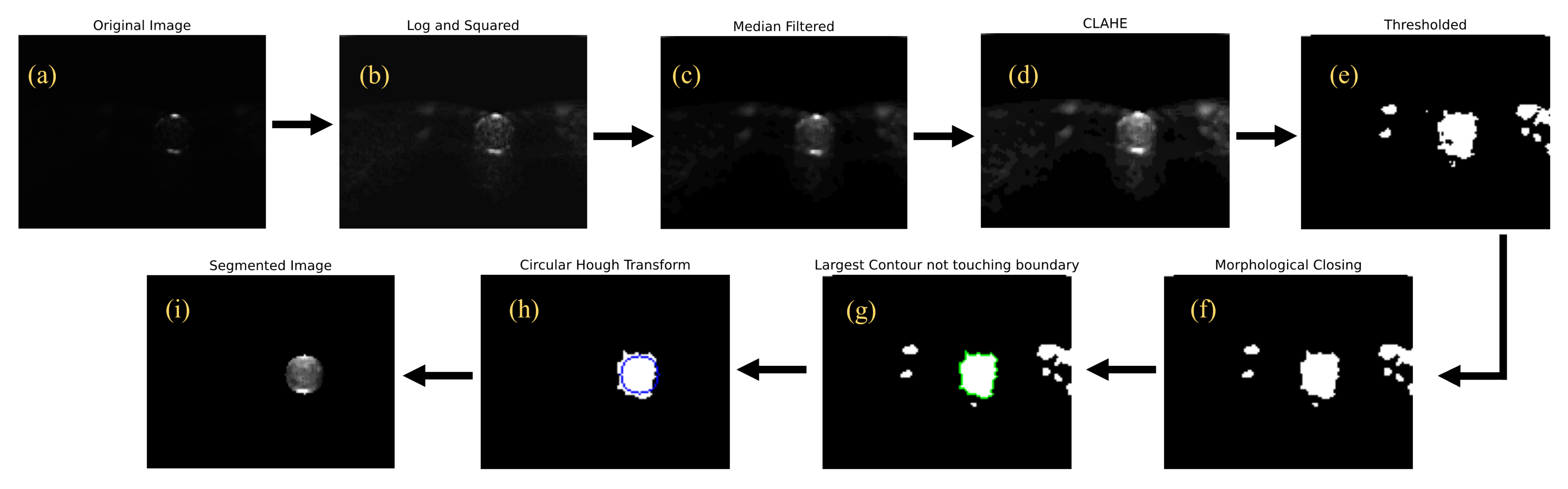}
        \caption{Image Processing Steps: (a) Acquired ultrasound Image, (b) After log compression and squaring, (c) Median filtered image, (d) Contrast limited adaptive histogram equalized image, (e) Thresholded image, (f) Image after morphological closing, (g) Largest contour not touching boundary identified (green boundary), (h) Circle fitting using hough transform on the identified contour (blue circle), (i) Segmented region of interest }
        \label{Image Processing Steps}
    \end{center}
\end{figure*}

Due to the log compression, the salt and pepper noise is increased and to suppress the same, a non-linear median filter was applied. A $3\times3$ overlapping window was employed for median filtering in this work, resulting in the filtered image shown in Fig. \ref{Image Processing Steps} (c).

\subsubsection{Contrast Limited Adaptive Histogram Equalization}

To further enhance the image quality, contrast limited adaptive histogram equalisation (CLAHE) \cite{c10}, a variant of adaptive histogram equalisation is applied, that distributes the part of the histogram that exceeds the clip limit equally across all histograms to limit the contrast amplification to reduce amplified noise as demonstrated in Fig. \ref{Image Processing Steps} (d).

\subsubsection{Thresholding}
The next step is to segment the relevant region of interest. The initial step is to convert the image to binary using global thresholding (using a threshold value of 49 here), and the binary image is shown in Fig. \ref{Image Processing Steps} (e).

\subsubsection{Morphological Operations}
The morphological closing operation is performed on the thresholded image using a cross-shaped structuring element of size $3\times3$ and the result is shown in Fig. \ref{Image Processing Steps} (f).


\subsubsection{Largest Contour Detection}
The contours in an image are lines that outline objects or features with similar intensity levels. The Suzuki algorithm \cite{c12}  is employed to detect all the contours present in the binary image. The detected contours are sorted based on the area, with the largest one not touching the image boundary chosen as the target contour, as shown in Fig. \ref{Image Processing Steps} (g).

\subsubsection{Hough Transform}


Since the objective of the tube phantom study is to extract objects which are cylindrical in nature, the Hough Transform \cite{c13} is employed to determine the best-fitting circle for the largest contour identified in each frame as shown in Fig. \ref{Image Processing Steps} (h).

 \subsubsection{Segmentation}
 The region enclosed by the identified circle is used as a binary mask and is then applied on the histogram equalised image to produce the final segmented image in Fig. \ref{Image Processing Steps} (i).

\subsection{3D Reconstruction and Visualisation}
After processing the data, the segmented ultrasound images of the cross-sectional view are obtained and will be employed for the volume reconstruction. The reconstruction was performed using the trilinear interpolation method. It increases the resolution of the dataset by estimating intensity values for voxels that were not present in the original data, resulting in improved quality and detail of the final volume, leading to more accurate diagnosis and reduced visual artefacts.  Since the acquisition is done at high frame rates with subsequent frames having high correlation, trilinear interpolation is a suitable approach for effectively reconstructing the 3D volumes in the proposed work. 

The reconstructed ultrasound volume dataset is visualized using the volume rendering approach available in the Volume Viewer tool in Matlab\textsuperscript{\textregistered}, providing the three-dimensional structure of the tube. The algorithm traces rays from the viewpoint to each voxel in the volume and uses the intensity values of corresponding 2D images to determine the voxel's colour and opacity.  This generates a realistic 3D representation of the ultrasound data, giving a more comprehensive understanding of the underlying structures.

\section{EXPERIMENTS AND RESULTS}
\label{EXPERIMENTS AND RESULTS}

\begin{figure*}[t!]
    \centering
    \includegraphics[width=.90\textwidth]{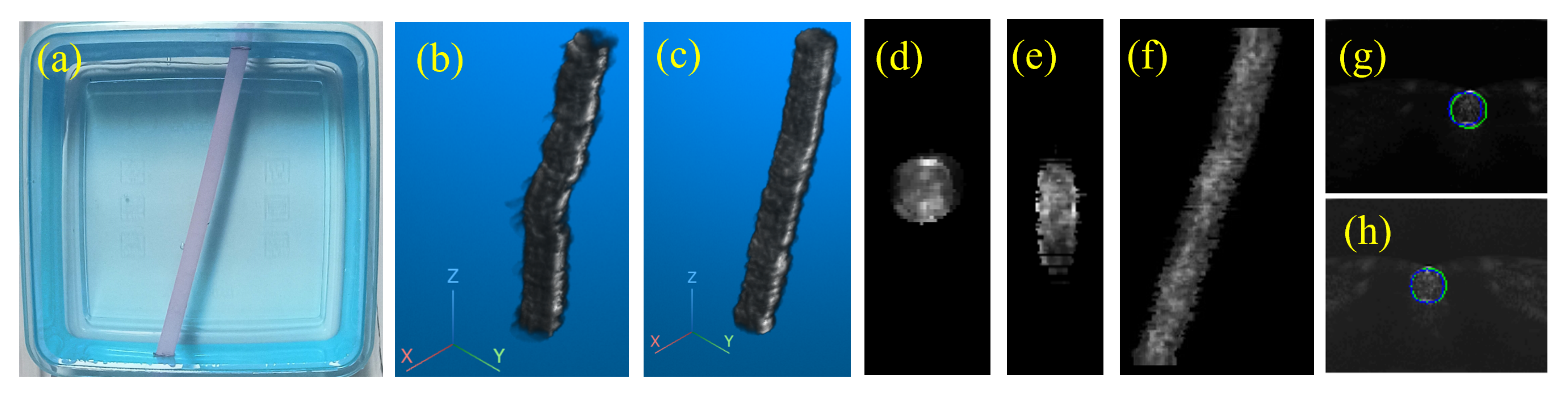}
    \caption{(a) Top view of the tube phantom. Reconstructed volume visualized in MATLAB: (b) data acquired without mechanical track, (c) data acquired with the mechanical track. (d) XY Slice, (e) YZ Slice, (f) XZ Slice, (g), and (h) Comparison of the auto segmentation (blue) and manual segmentation (green) for a random two slices.}
    \label{Reconstructed Volume along with ground truth}
\end{figure*}

\begin{figure}[t!]
    \centering
    \includegraphics[width=0.43\textwidth]{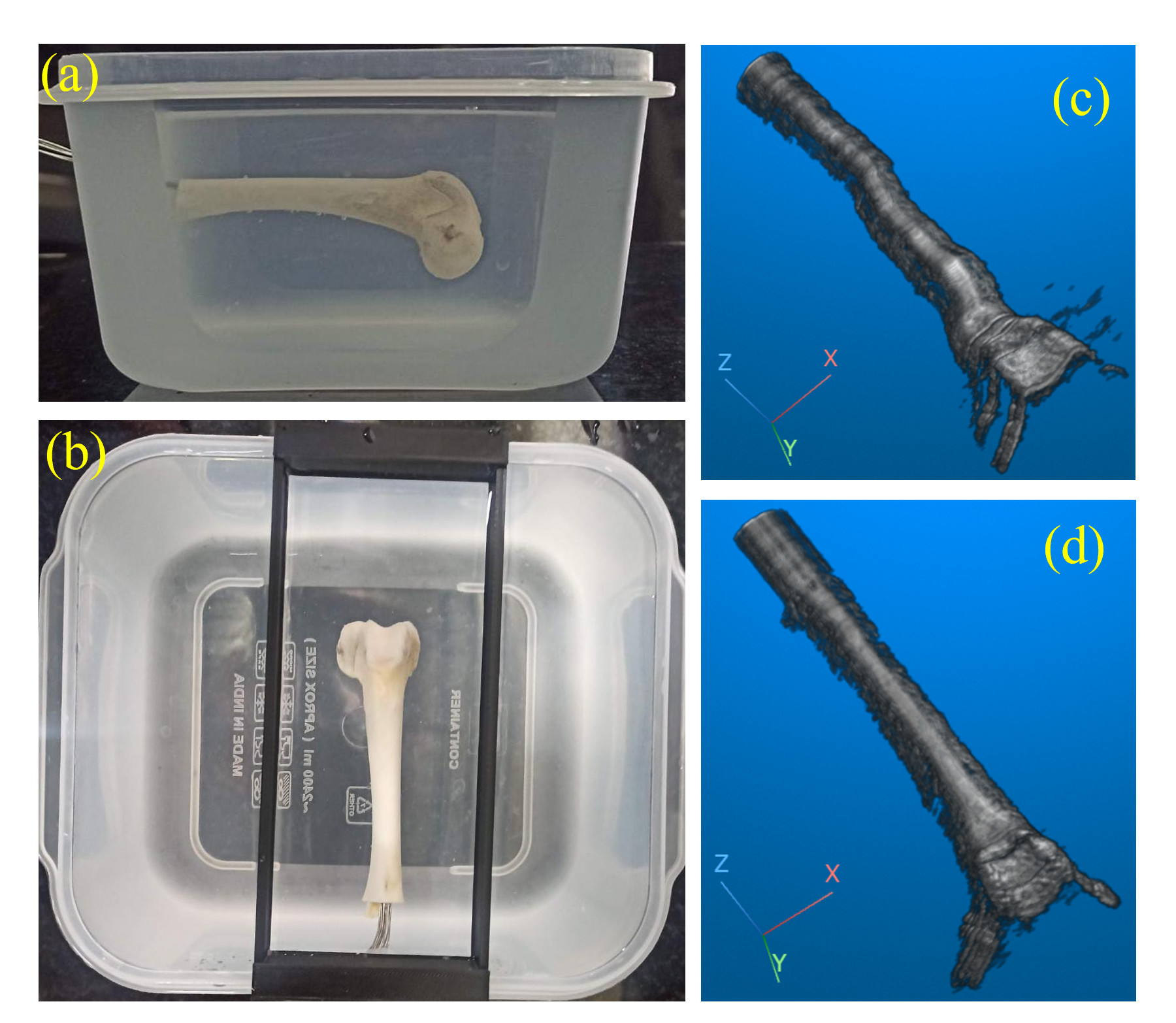}
    \caption{ \textit{Ex-vivo} bone phantom setup: (a) side view  (b) top view. Reconstructed volume visualized in MATLAB: (c) data acquired without the mechanical track (d) data acquired with the mechanical track.}
    \label{boneResults}
\end{figure}

\subsection{\textit{In-vitro} Tube Phantom Study}
The proposed approach was evaluated by reconstructing the 3D volume of a circular tube from the 2D ultrasound images acquired using the proposed mechanical track. A single-tube phantom was designed as shown in Fig. \ref{Reconstructed Volume along with ground truth} (a), and filled with a mixture of acoustic gel and chalk powder to act as a reflector. The tube was positioned in a cubic container filled with water to the top, ensuring that the tip of the probe touches water, thus ensuring impedance matching. The images were acquired using the 128-channel Verasonics Vantage research ultrasound platform with an L11-5v linear probe. Plane-wave transmission at a center frequency of $7.6\ MHz$ is employed for insonification and the delay and sum beamformer on the Verasonics vantage system is adopted for receive beamforming and image reconstruction. A total of $150$ beamformed grayscale images, each with dimensions of $100\times128$ pixels, were acquired with a single motion of the probe along the entire length of the mechanical track.

\AddToShipoutPictureBG*{%
  \AtPageUpperLeft{%
    \setlength\unitlength{1in}%
    \hspace*{\dimexpr0.5\paperwidth\relax}
    \makebox(0,-0.75)[c]{\textcolor{red}{\large This is an originally submitted version and has not been reviewed by independent peers.}}%
}}

\AddToShipoutPictureBG*{%
  \AtPageLowerLeft{%
    \setlength\unitlength{1in}%
    \hspace*{\dimexpr0.5\paperwidth\relax}
    \makebox(0,0.75)[c]{\textcolor{red}{\textit{This work is licensed under a \href{https://creativecommons.org/licenses/by-nc-nd/4.0/}{Creative Commons Attribution-NonCommercial-NoDerivatives (CC-BY-NC-ND) 4.0 License.}}}}%
}}

\subsection{3D Volume Visualisation and Evaluation}
 The reconstructed volume visualized in MATLAB is shown in Fig. \ref{Reconstructed Volume along with ground truth} (c), with $XY$, $YZ$ and $XZ$ slices displayed in Fig. \ref{Reconstructed Volume along with ground truth} (d), (e), and (f) respectively. The $XY$ slice represents the cross-sectional view of the tube, the tube was kept at a slant of $45^{\circ}$ so the $YZ$ slice gave an oval shape and the $XZ$ slice gives the top view of the phantom made.

The proposed auto segmentation is compared against the manual segmentation for 20 random samples from the image stack. The mean Intersection over Union (IoU) score between manual and automatic segmentation was determined to be 88.97\%. As seen in Fig. \ref{Reconstructed Volume along with ground truth} (g) and (h), the results demonstrate the high degree of similarity between manual and automatic segmentation.

\subsection{\textit{Ex-vivo} Validation}
The proposed approach is validated with an \textit{ex-vivo} experiment using a bone (tibia of a goat) placed in a water bath as shown in Fig. \ref{boneResults}(a) and (b). A total of $300$ beamformed grayscale images, each with dimensions of $128\times128$ pixels, were acquired with a single motion of the probe along the entire length of the mechanical track. The images are processed similarly to that of the \textit{in-vitro} study and the results are shown in Fig. \ref{boneResults}(d).

\section{Discussion and Conclusion }
\label{Conclusion and Discussion}
In this work, a novel framework for freehand scanning for 3D ultrasound imaging is proposed. The proposed approach consists of a 3D-printed mechanical track and a high frame rate acquisition. While the mechanical track ensured reduced errors caused by probe slippage or tilt while scanning, the high frame rate acquisition resulted in a high correlation among subsequent frames which helped in simpler approaches for volume reconstruction. The reconstructed volume of the \textit{in-vitro} tube phantom and the \textit{ex-vivo} bone phantom, shown in Fig. \ref{Reconstructed Volume along with ground truth}(b) and \ref{boneResults}(c) respectively, were generated using ultrasound images obtained without the use of a mechanical track. Comparing these volumes with that obtained using the mechanical track (Fig. \ref{Reconstructed Volume along with ground truth}(c) and Fig. \ref{boneResults}(d))  highlights the reduction in errors caused by probe slippage or tilt by incorporating the mechanical track in the imaging process. Detailed video results comparing 3D reconstruction with and without the use of the mechanical track are available online \cite{c25}. 

The proposed design is most effective for scanning on flat surfaces and can be improved by using a deformable material on the side that comes into contact with the scanning surface. To overcome errors due to variations in scanning velocity, a low-cost accelerometer can be employed for tracking probe motion and will be taken up as future work. While conventional segmentation techniques produced good results for the simple phantoms, neural network based segmentation techniques shall be employed for more complex structures as part of future work. Overall, the experimental results are encouraging and suggest that further research is warranted to address the current limitations to improve point-of-care ultrasound imaging.


\AddToShipoutPictureBG*{%
  \AtPageUpperLeft{%
    \setlength\unitlength{1in}%
    \hspace*{\dimexpr0.5\paperwidth\relax}
    \makebox(0,-0.75)[c]{\textcolor{red}{\large This is an originally submitted version and has not been reviewed by independent peers.}}%
}}

\AddToShipoutPictureBG*{%
  \AtPageLowerLeft{%
    \setlength\unitlength{1in}%
    \hspace*{\dimexpr0.5\paperwidth\relax}
    \makebox(0,0.75)[c]{\textcolor{red}{\textit{This work is licensed under a \href{https://creativecommons.org/licenses/by-nc-nd/4.0/}{Creative Commons Attribution-NonCommercial-NoDerivatives (CC-BY-NC-ND) 4.0 License.}}}}%
}}
\addtolength{\textheight}{-12cm}   






\end{document}